\documentclass{Pos}
\usepackage{amsmath}
\usepackage{amssymb}

\title{The unexpected role of D waves in low-energy neutral pion photoproduction}
\ShortTitle{The unexpected role of D waves in low-energy neutral pion photoproduction}

\author{\speaker{C\'esar Fern\'andez-Ram\'{\i}rez}\\
Center for Theoretical Physics, Laboratory for Nuclear Science and Department of Physics,
Massachusetts Institute of Technology,
77 Massachusetts Ave., Cambridge, MA  02139, USA\\
        E-mail: \email{CEFERA@MIT.EDU}}

\abstract{It has been commonly assumed that low-energy neutral pion photoproduction from the proton
can be described accounting only for S and P waves, and that higher partial waves are irrelevant.
We have
found that this assumption is not correct and that the inclusion of D waves is necessary to
obtain a reliable extraction of the $E_{0+}$ multipole from experimental data.
This is due in large measure to the spontaneous breaking of chiral symmetry in QCD
which leads to very small S-wave contributions. This makes the usual partial
wave expansion less accurate and although D waves are small, their contribution is
enhanced through the interference with P waves,  
which compromises the S-wave extraction from data if D waves are not taken into account.
In our work we have used
Heavy Baryon Chiral Perturbation Theory to one loop, and up to ${\cal O}(q^4)$,
to account for the S and P waves, while D waves are added 
in an almost model-independent way using standard Born
terms and vector mesons. We also show that higher partial waves do not play an important role.}

\FullConference{6th International Workshop on Chiral Dynamics\\
		 July 6-10 2009\\
		 Bern, Switzerland}

\begin{document}

\section{Introduction}

Due to the spontaneous breaking of chiral symmetry in Quantum Chromodynamics (QCD) the $\pi$ meson appears as a pseudoscalar Nambu--Goldstone boson \cite{book} which induces
some dynamical consequences. Among them, one of the most prominent is the softness of the
S-wave amplitude for the $\gamma N \rightarrow \pi^{0} N$ reaction in the near
threshold region, since it vanishes in the chiral limit \cite{CHPT91,CHPT96,CHPT01}. 
Accordingly,  the photoproduction of neutral pions differs from the general pattern 
for hadronic reactions where
the S wave dominates close to threshold and then, as the energy increases, the higher angular momentum waves (P, D, \ldots)  start to become important. By contrast, 
and also due to the large P-wave amplitude, for the 
$\gamma N \rightarrow \pi^{0} N$ reaction the S- and P-wave contributions
are comparable  even very close to threshold \cite{AB-fits}.
Hence, the accurate extraction of  S and P waves from pion photoproduction data becomes
an important issue in the study of the breaking of chiral symmetry and QCD.

It is important to recall that  the partial waves (electromagnetic multipoles) 
are not experimental observables, but rather are quantities extracted from data
using some kind of approach and/or theoretical input 
--- Heavy Baryon Chiral Perturbation Theory (HBCHPT), for instance.
So, the accuracy of the extraction relies on the quality of the approximations involved. 
In this Proceeding we expose the importance of including
D waves in the analysis even at the lowest pion production energies \cite{FBD09}.

\section{Impact of D waves in the extraction of the S wave  and the Low Energy Constants}

The standard approach to extract the S and P waves
consists in analyzing data assuming that these two partial waves are sufficient to
describe the experimental data, so that higher partial waves may be neglected
\cite{CHPT91,CHPT96,CHPT01,AB-fits,Schmidt,AB-review}.
This assumption has been sustained in previous analyses by two arguments:
\begin{enumerate}
\item The angular dependence of the experimental differential cross section
can be described accurately using  Legendre polynomials up only to order two.
Indeed, S and P waves constitute the minimal set of partial waves needed to reach that angular dependence;
\item In the near-threshold energy region higher partial waves are weak
and the early dominance of the $M_{1+}$ multipole renders them negligible.
\end{enumerate}

Let us discuss these two statements. The differential cross section expanded in terms of Legendre polynomials and up to D waves reads
\begin{equation}
\frac{d\sigma}{d\Omega} = \frac{q_\pi}{k_\gamma} \left[ T_0 
+ T_1 \mathcal{P}_1\left( \theta \right)  
+ T_2 \mathcal{P}_2\left( \theta \right)
+ T_3 \mathcal{P}_3\left( \theta \right) 
+ T_4 \mathcal{P}_4\left( \theta \right) \right] \: , 
\label{eq:sigma}
\end{equation}
where $q_\pi$ and $k_\gamma$ are the pion and photon momenta in the center-of-mass, respectively,
and the $T_i$ depend on the photon energy. If we restrict ourselves to up to P waves, then $T_3=T_4=0$, two coefficients that are not necessary to describe current experimental data, backing the first statement. However, the first argument is misleading because any coefficient that accompanies a Legendre polynomial in the expansion depends on \textit{all} partial waves,
implying that in some circumstances 
higher partial waves can make an important contribution to the coefficients
that accompany the lower-order Legendre polynomials, 
thereby posing uncertainties in the multipole extraction.
This assumption cannot be taken for granted and has to be tested.
The second argument is actually correct for $T_0$ and $T_2$ because they contain the dominant $|M_{1+}|^2$ contribution which makes any D wave content negligible. However, in the case of $T_1$ the situation is quite different and the argument is in favor of the
the possible importance of higher partial waves. $T_1$ reads \cite{FBD09,FBD_PRC09}
\begin{equation}
T_1= 2 \: \text{Re} \left[ P_1^* E_{0+} \right] + \delta T_1 \: ,
\end{equation}
where $P_1\equiv 3E_{1+}+M_{1+}-M_{1-}$, 
and $\delta T_1$ stands for the D-wave/P-wave interference contribution
\begin{equation}
\begin{split}
\delta T_1 =& 2 \: \text{Re} \Big{[} \: \frac{27}{5}M^*_{1+}M_{2+} 
+\left( M^*_{1+} - M^*_{1-} \right) E_{2-} 
+ \left( \frac{3}{5}M^*_{1+}+3M^*_{1-} \right) M_{2-} \\
&+E^*_{1+} \left( \frac{72}{5}E_{2+}-\frac{3}{5}E_{2-} +\frac{9}{5}M_{2+}
-\frac{9}{5}M_{2-} \right)  \Big{]} \: .
\end{split}
\end{equation}
The conclusion disregards that the S wave is also weak
and that important features of the angular dependence of the observables
are dominated by the interference of different partial waves. In this situation the dominance of a certain contribution (such the $M_{1+}$ in this case) can lead to an important enhancement of smaller partial waves through interference, making them relevant. 
In order to understand the interplay of the electromagnetic multipoles and the impact of the D waves
one can work out a pedagogical approximation of $T_1$. Let us assume that $E_{1+}$ and $M_{1-}$
are negligible. In other words, that $M_{1+}$ is much larger than the other two P waves. If it is so, $T_1$ can be written
\begin{equation}
T_1 \simeq 2 \: \text{Re} \left[ \:
M^*_{1+} \left( E_{0+} +\frac{27}{5}M_{2+} + E_{2-}+  \frac{3}{5}M_{2-} \right)  \right] \: .
\end{equation}
If D waves are not included explicitly, the extracted $E_{0+}$ has a certain D-wave {\it pollution} that has to be calculated. In a back of the envelope calculation one can find easily that above the $\pi^+$ production threshold $\frac{27}{5}M_{2+}$ is of the order of a 6\% of the $E_{0+}$. 
Hence, a full calculation including D waves becomes mandatory.

As previously said, a certain theoretical framework has to be set to extract the electromagnetic multipoles. In our case, we have chosen HBCHPT to compute the S and P waves, 
which constitutes the best available theoretical framework to study pion photoproduction in the near threshold region. In \cite{CHPT96,CHPT01} the S and P multipoles to one loop
and up to $\mathcal{ O}(q^4)$ are provided and we take these as our starting point.
The D waves are computed using 
the customary Born terms (equivalent to the Born contribution to HBCHPT)
and vector meson exchange ($\omega$ and $\rho$) 
\cite{FMU06}.
For the $\omega$ and $\rho$ parameters we have used
those given by the dispersion analysis of the form factors in \cite{MMD}.
The vector-meson correction is very small and the inclusion of D waves in this way is almost 
equivalent to the zeroth order in HBCHPT.
The inclusion of vector mesons entails a certain model dependency, however, this dependency is very small and the results we obtain regarding the impact of D waves are effectively model independent.

\begin{figure}
\rotatebox{-90}{\includegraphics[width=.35\textwidth]{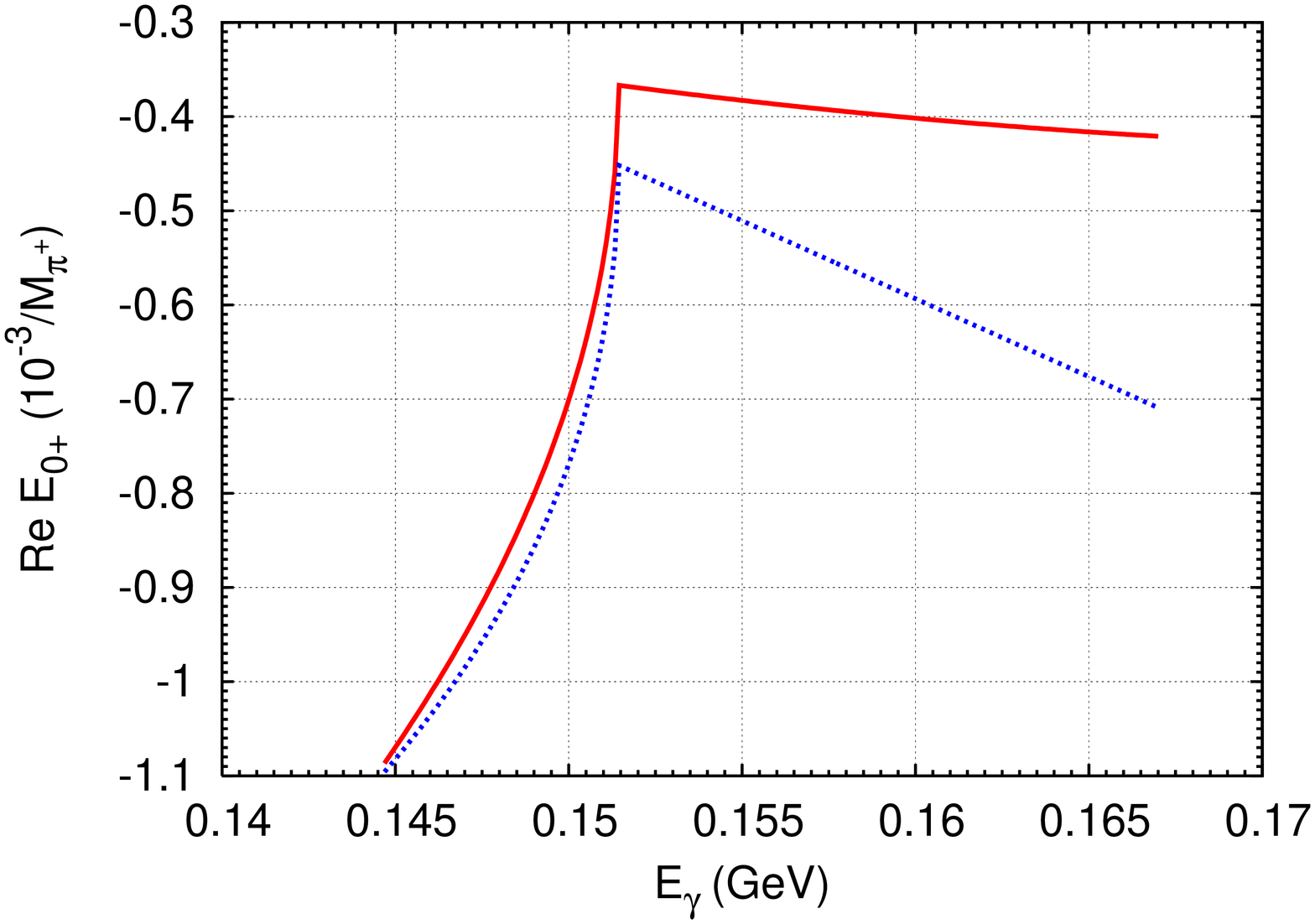}}
\rotatebox{-90}{\includegraphics[width=.35\textwidth]{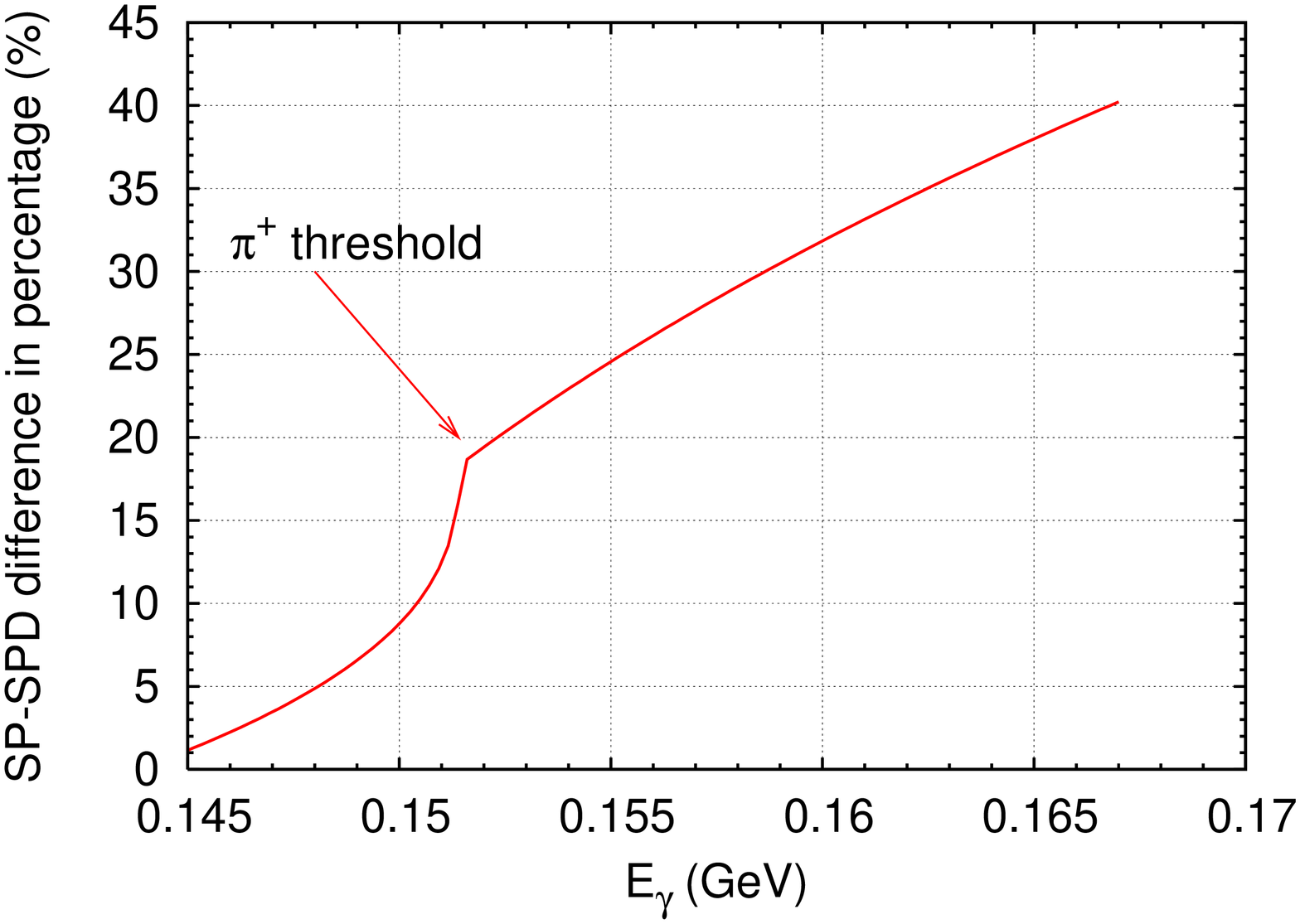}}
\caption{Left panel: Extracted $E_{0+}$ multipole. The solid line corresponds to the SPD calculation and the dotted line to the SP calculation. Right panel: Extracted $\text{Re} E_{0+}$ multipole ratio between the SP and SPD fits.}
\label{fig1}
\end{figure}

Within this framework we have performed fits
to the latest experimental data from MAMI \cite{Schmidt} 
(171 differential cross sections and 7 photon asymmetries, spanning the energy range from threshold up to 166 MeV) using either solely S and P waves 
(SP fit) or  S, P and D waves (SPD fit).
As a fitting procedure
we have used a hybrid optimization code based on a genetic algorithm \cite{PRC08}.

Both the SP and SPD calculations provide equally good fits to data from \cite{Schmidt} and the same P waves within a 1\% level, but the extracted S waves are rather different.
In Fig. \ref{fig1} we provide the resulting extraction of the $E_{0+}$ multipole.
In the left panel we compare both SP and SPD extractions and in the right panel we provide the difference between them in percentage. It is noticeable that as early as at the charge pion threshold the impact of D waves is nearly a 20\% effect.

\begin{figure}
\begin{center}
\rotatebox{0}{\includegraphics[width=.5\textwidth]{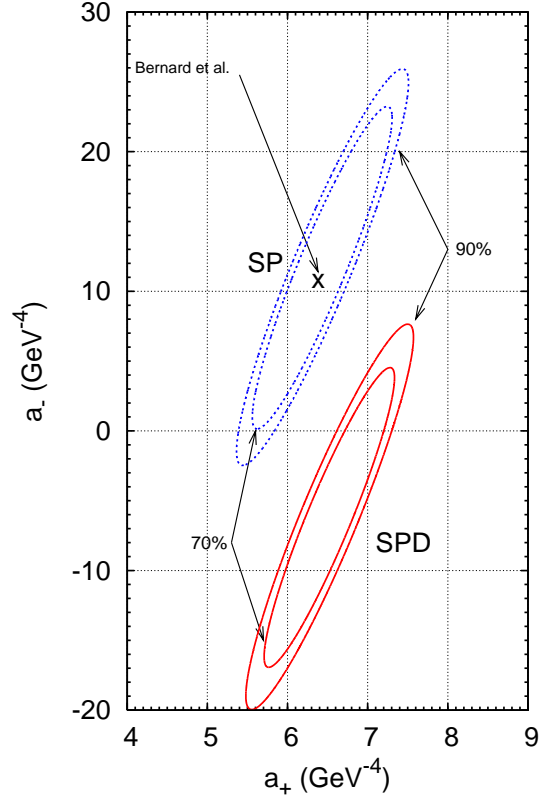}}
\end{center}
\caption{Correlation plot for $a_+$ and $a_-$ given the P-wave parameters. 
We display both the SP (dotted) and SPD (solid) fits.
The cross is the value of $a_+$ and $a_-$ given by Bernard \textit{et al.}  \cite{CHPT01}.
For each fit, the inner line is the 70\% confidence level and the outer one is the 90\% line.
Note that the different scales have been used for $a_+$ and $a_-$.}
\label{fig2}
\end{figure}

Because we are employing HBCHPT to describe the $E_{0+}$ multipole, every difference between both extractions is embedded in the Low Energy Constants (LECs) $a_1$ and $a_2$ associated to the 
$E_{0+}$ counter-term.
As noticed by Bernard, Kaiser, and Mei\ss ner \cite{CHPT01}, $a_1$ and $a_2$ are two highly
correlated LECs, and their sum is a better quantity to use when analyzing data. The reason is that 
the leading order for the counter-term is the one associated with $a_+=a_1+a_2$, namely the parameter
that accounts for the value of $E_{0+}$ at the neutral pion threshold:
\begin{eqnarray}
\text{SP fit:} &\:&    a_+ = 6.48^{+0.82}_{-0.93} \: \text{\small (70\% C.L.)} ^{+1.07}_{-1.13}  \: \text{\small  (90\%C.L.)} \: , \\
\text{SPD fit:} &\:& a_+ = 6.57^{+0.83}_{-0.92} \: \text{\small (70\% C.L.)} ^{+1.03}_{-1.12}  \: \text{\small  (90\% C.L.)} \: .
\end{eqnarray}
For $a_-=a_1-a_2$ the uncertainties are too large and, if any good extraction of $a_-$ is intended, the inclusion of D waves is very important. In Fig. \ref{fig2} we show the correlation plot for the $(a_+,a_-)$ pair. The big error bars associated to $a_-$ are apparent.

\section{Conclusions}
The main results we have obtained can be summarized in:
\begin{enumerate}
\item S and P waves alone are not enough to analyze the experimental data and extract accurately the electromagnetic multipoles. Without the inclusion of D waves the extraction of the S wave is compromised and so are the LECs of the $E_{0+}$ multipole and the conclusions that can be derived. 
Hence, D waves cannot be dismissed in the analysis of low-energy pion photoproduction.
\item The extraction of the P waves is not affected by the inclusion of D waves at the 1\% level.
\item D waves do not impact the value of $E_{0+}$ at the neutral pion threshold --- this is made apparent by the fact that both the SP and SPD fits yield very close $a_+$ values ---, but they have an almost 20\% impact at the charged pion threshold, an effect that increases with energy.
\item The impact of D waves in low-energy pion photoproduction is due to the soft nature of the S wave
and a direct consequence of chiral symmetry and the Nambu-Goldstone nature of the pion.
\end{enumerate}

At this point one might wonder if F waves could also make an important contribution to $T_1$
in this energy region through their interference with P waves.
The answer is no because, due to symmetry considerations, there is no interference of F and P waves
in $T_1$. Indeed, F waves only interfere with D and G waves, guaranteeing a negligible contribution
to $T_1$ in the near-threshold region. A detailed analysis on the impact of F and higher partial waves in  observables can be found in \cite{FBD_PRC09}, where it is proved that they can be neglected.

In the coming years our knowledge of the electromagnetic multipoles will be increased thanks to the
forthcoming  \cite{MAMI-Bernstein} and currently under analysis \cite{Hornidge} experiments at MAMI.
The cross section and $\Sigma$ asymmetry data at several energies \cite{Hornidge}  will allow one to pin down the P waves with higher accuracy, and the $T$ and $F$ 
asymmetries \cite{MAMI-Bernstein} will allow one to
extract both the real and the imaginary parts of the S wave.
In addition the measurement of the $E$ asymmetry would serve to 
unambiguously determine the D-wave contribution \cite{FBD_PRC09}. 

\acknowledgments

The work reported here was carried out in collaboration with Aron M. Bernstein and T. William Donnelly.
This research was supported in part by 
"Programa Nacional de Movilidad de Recursos Humanos del Plan Nacional I+D+I 2008-2011"
of Ministerio de Ciencia e Innovaci\'on (Spain).
We gratefully acknowledge the Institute for Theoretical Physics at Bern University for their hospitality.

\end{document}